\documentclass[aps,pre,twocolumn,showpacs,floatfix,superscriptaddress]{revtex4}
\usepackage{graphicx}
\usepackage{bm}
\bibstyle{apsrev.bib}

\begin{document}
\title{Universality in the pair contact process with diffusion}
\author{G. T. Barkema}
\affiliation{Institute for Theoretical Physics, Utrecht University,
3584 CE Utrecht, The Netherlands}
\author{E. Carlon}
\affiliation{Theoretische Physik, Universit\"at des Saarlandes,
D-66041 Saarbr\"ucken, Germany}

\date{\today}

\begin{abstract}
The pair contact process with diffusion is studied by means of
multispin Monte Carlo simulations and density matrix renormalization
group calculations. Effective critical exponents are found to behave
nonmonotonically as functions of time or of system length and extrapolate
asymptotically towards values consistent with the directed percolation
universality class.  We argue that an intermediate regime exists where the
effective critical dynamics resembles that of a parity conserving process.
\end{abstract}

\pacs{05.70.Ln, 64.60.Ak, 64.60.Ht}

\maketitle

\section{Introduction}

Out of equilibrium systems may display phase transitions analogous to
those found in their equilibrium counterparts. These transitions are
classified into distinct universality classes, characterized by a set of
critical exponents \cite{hinr00,marr96}.  Particular attention has been
paid to one-dimensional systems with transitions from an active state
into absorbing states, i.e., frozen configurations from which the system
cannot escape. For these systems, so far, only two distinct universality
classes have been firmly established: the directed percolation (DP)
and the parity conserving (PC) universality class. While the former is
ubiquitous and found in several systems with very different dynamical
rules, the latter is only known to occur when extra symmetries are
present \cite{hinr00,marr96}.

A model which has attracted quite some interest recently, because
it may indeed belong to a novel universality class \cite{howa97}
is the so-called pair contact process with diffusion (PCPD).
Despite a rather intense activity in the past couple of years
\cite{carl01,hinr01,odor00,noh01,dick02,park02,hinr02a,kock03,odor03},
the understanding of the PCPD and its relation with other known models
is still unsatisfactory. In the {\it fermionic} version of the 
model---the one studied here---each lattice site is either occupied by a
single particle ($A$) or empty ($0$). The reactions are
\begin{eqnarray} 
\begin{array}{ccc}
\left\{ 
\begin{array}{ccc} 
AA0 & \rightarrow & AAA \\ 
0AA & \rightarrow & AAA 
\end{array} \right.  
&{\rm with \,\,\, rate}&
\frac{(1-p)(1-d)} 2, \label{2Ato3A}\\ 
AA~ \rightarrow ~ 00
&{\rm with \,\,\,rate}& p
(1-d), \label{2Ato0}\\ 
A0  \leftrightarrow  0A 
&{\rm with \,\,\,rate}&  d 
\label{A0to0A}
\end{array} 
\label{eq1}
\end{eqnarray} 
with $0 < d < 1$, $0 < p <1$.  The analysis of the critical properties
of the PCPD has shown to be much more difficult than all similar models
analyzed so far and several scenarios have been proposed. First, a
similarity of the exponents $\beta/\nu_\perp$ and $z=\nu_\|/\nu_\perp$
(where $\nu_\|$ and $\nu_\perp$ are the correlation length exponents along
the time and space directions and $\beta$ the order parameter exponent)
with those of the PC class \cite{carl01} was reported, although in the
PCPD there is no conservation of parity.  It was later suggested that the
PCPD could belong to a new universality class with exponents close to,
but different from the PC values \cite{hinr01}.  It has also been argued
that there could be two universality classes \cite{odor00} at small
and large diffusion rates ($d$), or continuously varying exponents
\cite{noh01}, or that scaling may even be violated \cite{dick02}.
More recently the option of a slow crossover to DP was also discussed
\cite{hinr02a}, although the general belief is that the PCPD belongs
to a novel universality class \cite{noh01,dick02,park02,kock03,odor03}.
Field theoretical methods, which have been successfully applied to other
reaction--diffusion models \cite{card96} failed so far to clarify the
critical properties of the PCPD \cite{howa97,muno02}.

In this paper we present some insights into the PCPD. We show that
accurate numerical results from Monte Carlo (MC) simulations and
density matrix renormalization group (DMRG) calculations convincingly
demonstrate that for sufficiently long times and system lengths the
exponents crossover towards the DP values.  Corrections to scaling
are, however, rather strong and only an accurate extrapolation of the
effective critical exponents allows to identify the final asymptotic
critical behavior.  We give evidence that in the near-asymptotic region
the model shows effective exponents close to the PC class values, which
suggests that the critical behavior of the system is described by two
competing fixed points.

\section{Results on bulk particles and pair densities}

Our MC simulations exploit a technique known as multispin coding
\cite{newm99}.  The basic idea is that in a simulation of 64 systems, each
with $L$ sites, the occupation of site $i$ in the $k${th} simulation is
stored in the $k${th} bit of 64-bit word $A[i]$. To perform the reaction
$AA0 \rightarrow AAA$ in all 64 systems at a randomly chosen site $i$
and its neighbors, one logical operation $A[i+1]=A[i+1]\lor (A[i] \land
A[i-1])$ suffices, where $\lor$ and $\land$ are the logical operations
OR and AND, respectively. Other reactions require only slightly more
elaborate logical operations.  A direct implementation along these lines
might result in 64 simulations, each of which statistically correct,
but strongly correlated to each other because the site selection is
shared. To alleviate this correlation without sacrificing efficiency, we
employ random bit patterns that decide which reaction will be attempted in
which system. The strong point of multispin coding is its efficiency. As
illustrated above, only a few logical operations (each usually carried
out in a single clock cycle without delay) suffice to update a site
in 64 systems simultaneously. For each combination of $d$ and $p$ we
simulated 64 systems with $L=100\ 000$ sites over $3\times 10^6$ Monte
Carlo time units, in about 15 hours on a single--processor workstation.
We also simulated $16\times 64$ systems with $L=100\ 000$ sites over
$10^7$ Monte Carlo time units on a parallel computer, for $d=0.5$ and
$p=0.152\ 45$, our estimate for the critical point.

\begin{figure}[t]
\includegraphics[width=7.7cm]{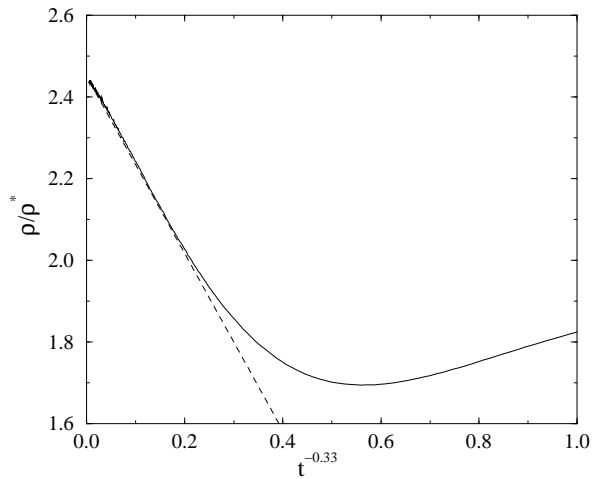} 
\caption{Ratio of the particle density $\rho$ and pair density $\rho^*$,
as a function of $t^{-\gamma'}$, with $\gamma'=0.33$, for the PCPD with
$d=0.5$ and at the estimated critical point $p=0.152\ 45$. 
This ratio approaches a constant.} 
\label{FIG01} 
\end{figure} 

A standard procedure to obtain critical exponents by means of
Monte Carlo simulations is to study the decay of the particle
density $\rho(t)$ as a function of time $t$, starting from a random
configuration of particles. At the critical point one has $\rho(t)
\sim t^{-\beta/\nu_\|}$. To monitor the decay it is convenient to
define the effective exponent $\delta_{\rm eff} \equiv - \partial \ln
\rho(t)/\partial \ln t$.  Typically $\delta_{\rm eff}$ is plotted as a
function of $1/t$. At the critical point, in the limit $t \to \infty$,
it approaches a finite value ($\delta_{\rm eff} \to \beta/\nu_\|$), while it
deviates upwards or downwards with respect to this value in the inactive
and active phases, respectively. This criterion allows to estimate the
critical point location and the ratio $\beta/\nu_\|$ \cite{hinr00}.
Some care, however, has to be taken when corrections to scaling are
particularly strong, for example when $\rho(t) \sim t^{-\beta/\nu_\|} \left( 1 +
c t^{-\gamma} \right)$ with $\gamma < 1$ and $c$ a constant. In this
case $\delta_{\rm eff}$ plotted as a function of $1/t$ approaches
$\beta/\nu_\|$ with an infinite slope. The ideal situation would be to
plot $\delta_{\rm eff}$ as a function of $1/t^\gamma$ as in that case the
approach to the asymptotic value would be linear. Further on in this paper
we will give numerical evidence that the correction-to-scaling exponent
$\gamma$ is close to $\beta/\nu_{\|}$ (which is much smaller than 1). A
natural choice is therefore to plot $\delta_{\rm eff}$ as a function of
the particle density $\rho$ instead of $1/t$, to avoid infinite slopes.

Besides the decay of the particle density $\rho$, we also consider
the decay of the pair density $\rho^* \equiv \langle A A \rangle$.
Before presenting the results for the critical exponents we analyze
the behavior of the ratio $\rho / \rho^*$ at the critical point.
Such a quantity, which is shown in Fig. \ref{FIG01} for $d=0.5$,
approaches a constant value ($\approx 2.45$) asymptotically for long
times. The most important consequence of this fact is that in PCPD one
can extract the critical exponent $\beta/\nu_\|$ both from the decay of
the particle and pair densities \cite{ratio}. Numerically the ratio is
much better behaved than the individual densities $\rho$ and $\rho^*$,
as fluctuations in the individual densities are highly correlated and
largely cancel each other in the ratio.

\begin{figure}[t]
\includegraphics[width=7.7cm]{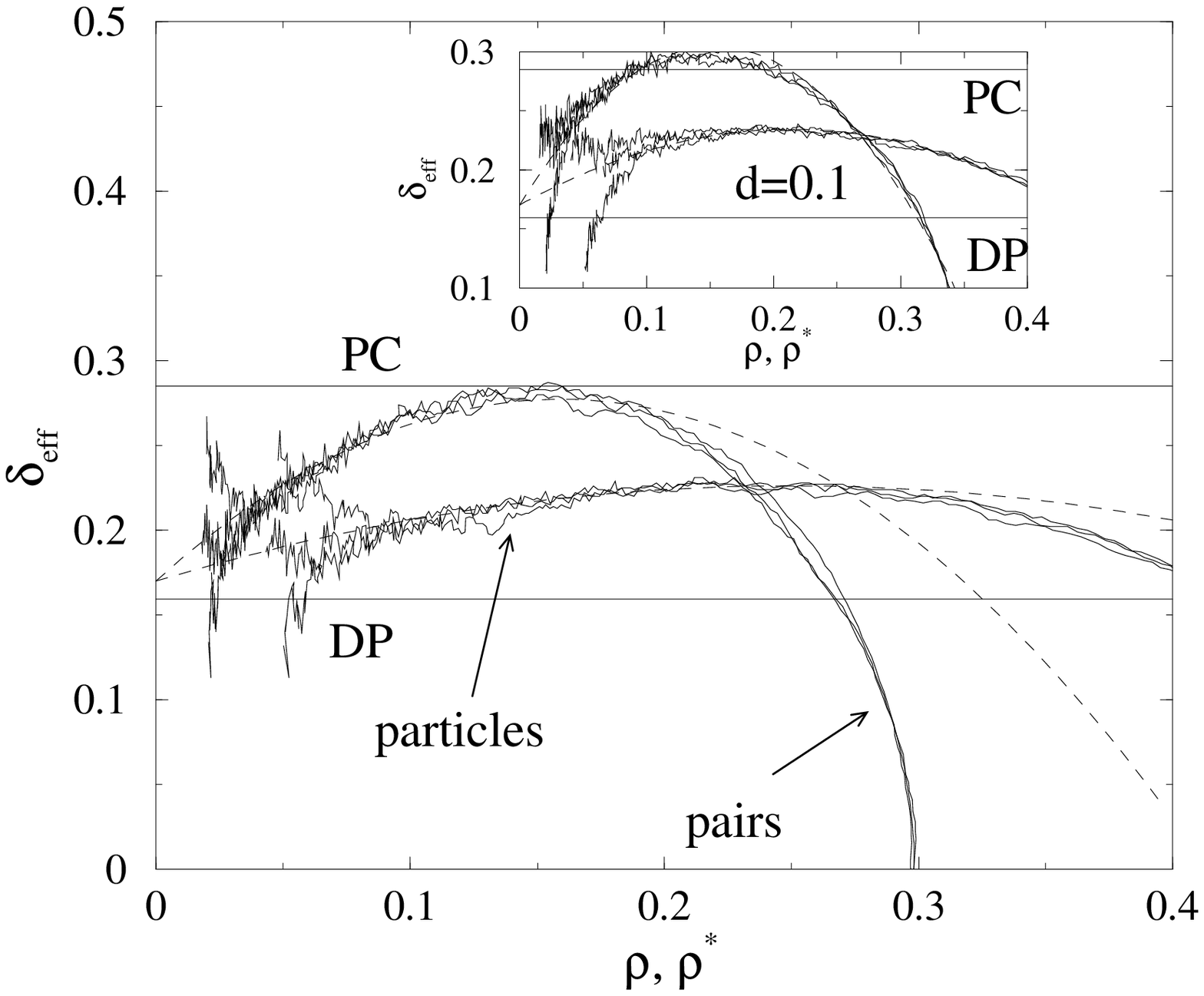}
\caption{Plot of $\delta_{\rm eff}$ at $d = 0.5$ and $p=0.1524$,
$p=0.152\ 45$, and $p=0.1525$ (from bottom to top) obtained from the decay of
the particles and pairs densities. The dashed lines are parabolic fits to
the data. The two horizontal lines show the ratio $\beta/\nu_\|$ for the
DP and PC class. Inset: $d=0.1$ and $p=0.1110$, $p=0.111\ 05$, $p=0.1111$.}
\label{FIG02}
\end{figure}

The data of Fig. \ref{FIG01} also provide an estimate of the leading
correction term in the asymptotic limit $t \to \infty$, which appears to
be of the type $\rho / \rho^* \sim C (1 + D t^{- \gamma'})$ with $\gamma'
\approx 0.33$, as shown by the linear approach to the asymptotic behavior
of the data when plotted versus $t^{- 0.33}$.  This exponent is much
larger than its equivalent $\gamma$ in the particle and pair densities,
which implies that the leading corrections for $\rho$ and $\rho^*$ cancel
in the ratio $\rho / \rho^*$. We tested that a similar cancellation
also occurs in the process $A \to 3A$, $2A \to 0$, which belongs to the
PC universality class \cite{hinr00}: the leading correction in $\rho$
and $\rho^*$ scales as $t^{-0.6}$, whereas the leading correction in
the ratio $\rho / \rho^*$ scales as $1/t$.

Figure \ref{FIG02} shows a plot of $\delta_{\rm eff}$ at $d=0.5$ as a
function of $\rho$ and $\rho^*$, calculated in the PCPD from the decay
of particles and pairs, respectively. Three different values for $p$
have been plotted around the critical point which we estimate as $p_c
= 0.152\ 45(5)$.  For $p > p_c$ (inactive phase) and $p < p_c$ (active
phase) $\delta_{\rm eff}$ rapidly veers up and down, as expected.  At the
critical point, $\delta_{\rm eff}$ approaches the $y$ axis with a finite
slope, indicating that the leading correction to scaling is most likely
described by an exponent roughly equal to $\delta$ itself.  A parabolic
fit through the data yields as a common estimate for the critical exponent
$\beta/\nu_\| = 0.17$. Similar calculations were repeated for other values
of the diffusion coefficient $d=0.05$, $0.1$, $0.2$, and $0.9$ (the inset
of Fig. \ref{FIG02} shows the case $d=0.1$), with the same results which
we can summarize with the estimate $\beta/\nu_\| = 0.17(1)$. This value
is consistent with the DP class exponent $\beta/\nu_\| = 0.159$.

\begin{figure}[t]
\includegraphics[width=7.7cm]{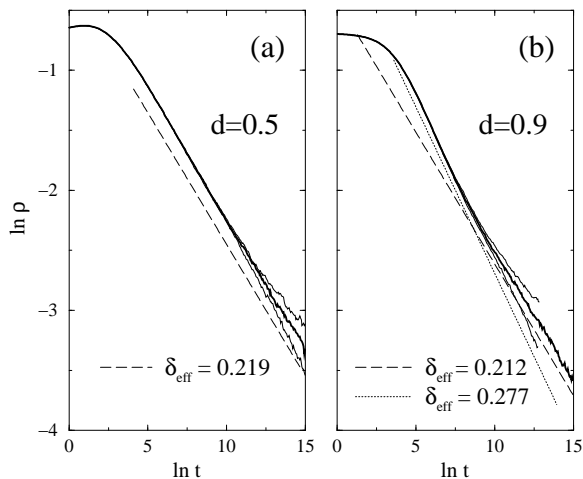} 
\caption{Log-log plots of $\rho$ (particle density) vs $t$. (a) $d=0.5$,
$p=0.1524$, $p=0.152\ 45$ and $p=0.1525$ (from top to bottom). (b) $d=0.9$
and $p=0.2330$, $p=0.2335$ and $p=0.234$ (from top to bottom). The
critical point densities are plotted as thick solid lines. The dashed
lines are linear fits to the data, shifted for clarity.}
\label{FIG03} 
\end{figure} 

An alternative way to calculate critical exponents would be to use a
linear fit of the densities versus time on a double-logarithmic scale.
However, such fits are hazardous when dealing with very slow convergence,
as is the case here, and may lead to wrong estimates for the critical
exponents.  We illustrate this in Fig. \ref{FIG03} which shows the
plot of the average particle density as a function of the time in a
double-logarithmic scale for $d=0.5$ and $d=0.9$. In the former case a
straight line (dashed) fits extremely well the critical density decay
leading to the estimate $\delta_{\rm eff} = 0.219$.  The analysis of the
effective exponent (see Fig. \ref{FIG02}), however, provides a closer
inspection of the local slopes of the double-logarithmic data. This
analysis reveals some remaining curvature, and the final estimate of
the exponent is significantly lower, compared to that obtained from the
fit in the double-logarithmic scale.  In the case of higher diffusion
[see Fig. \ref{FIG03}(b)] the curvature is more pronounced and clearly
visible also in the double-logarithmic plot, which can be fitted by two
straight lines with slopes $\delta_{\rm eff} \approx 0.277$ in the range
$4 \lesssim \ln t \lesssim 8$, and with $\delta_{\rm eff} \approx 0.212$
for $10 \lesssim \ln t \lesssim 15$.  Notice that the former exponent
is consistent with that expected for the PC class ($\delta_{\rm PC} =
0.286$ \cite{hinr00}).

Again an extrapolation of the effective exponent (as done in
Fig. \ref{FIG02}) shows convergence to a value consistent with DP. Note
that the value $\beta/\nu_\| \approx 0.21$ is consistent with the most
recent Monte Carlo estimates for the PCPD \cite{odor02,kock03,hinr02a}.
In particular, Kockelkoren and Chat\'e, \cite{kock03} performed a
series of Monte Carlo simulations for a bosonic version of the PCPD
where the constraint of one particle per site is released. Their
estimation of critical exponents is based on a straight-line fit to
a double-logarithmic plot of $\rho$ versus $t$, from which they find
$\beta/\nu_\| = 0.200(5)$. This bosonic version of the model is claimed
to suffer less from corrections to scaling than the fermionic case.
Notice, however, that also in the fermionic PCPD studied here the
density decay at $d=0.5$ (see Fig. \ref{FIG03}(a)) is rather straight
in a double-logarithmic plot for simulation times similar to those in
Ref. \cite{kock03}.  The advantage of the effective exponent analysis
performed here is that it allows to extrapolate the numerical results
to time scales {\it beyond} those actually simulated.

\begin{figure}[t]
\includegraphics[width=7.7cm]{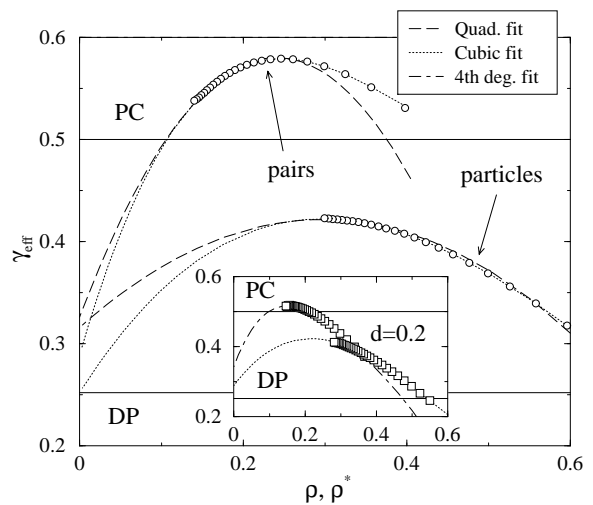}
\caption{
Plots of $\gamma_{\rm eff}$ for $d=0.5$ at the system critical point
($p=0.152\ 45$) for lattices up to $L=60$ calculated from the decay of
particle (circles) and pair (squares) densities as function of the
lattice lengths.  The dotted and dashed lines are fits in powers of
the densities. Inset: $\gamma_{\rm eff}$ for $d=0.2$.}
\label{FIG04} 
\end{figure}

Next we present some DMRG results. DMRG \cite{karen} allows to calculate
accurate stationary state probabilities for chains of moderate lengths
\cite{carl99}. As usual in DMRG, we used open boundary conditions. In
the PCPD on a lattice of finite length there are only two stationary
states: a state with no particles and a state occupied by a single
diffusing particle. To induce a finite density of particles we added
a reaction $0 \to A$ at the two boundary sites. The particle density
decays from the two boundaries and forms a U-shaped profile. For chains
of various lengths we calculated the density of particles $\rho(L)$
and of pairs $\rho^*(L)$ at the central site of a system of length $L$.
At the critical point these quantities decay in the limit $L \to \infty$
as $\rho(L) \sim \rho^*(L) \sim L^{-\beta/\nu_\perp}$.

Figure \ref{FIG04} shows the effective exponent $\gamma_{\rm eff} =
- \partial \ln \rho(L)/\partial \ln L$ versus $\rho$ for $d=0.5$ at the
critical point. As in Fig. \ref{FIG02} we include also the data
for the pairs. Dotted and dashed lines are fits with polynomials in
the densities. Again, a test of good convergence is that both exponents
extrapolate to the same asymptotic value. This requirement seems indeed
to be fulfilled and we find as extrapolation $\beta/\nu_\perp = 0.27(4)$.
This exponent is again consistent with the DP value $\beta/\nu_\perp =
0.252$ \cite{hinr00}. Similar results have also been found for other
values of the diffusion coefficient $d$. Extrapolations for $d=0.2$
are shown in the inset of Fig. \ref{FIG04}. At small $d$ the maxima in
$\gamma_{\rm eff}$ shift to longer $L$, thus extrapolations are somewhat
less stable. In this case we take the estimate obtained from the particles
$\beta/\nu_\perp = 0.28(5)$.

Previous DMRG results \cite{carl01} were restricted to the density of
particles and to smaller systems than that studied here. The data for the
effective exponent showed a monotonic behavior (except at very strong
diffusion) and were analyzed using an extrapolation with polynomes in
$1/L$. These extrapolations lead to a value consistent with the PC class
exponent $\beta/\nu_\perp = 0.50$ \cite{carl01}.  The present calculation,
extended to the density of pairs and to longer systems, reveals that
nonmonotonicity in the effective exponent is a common feature at all
$d$. This nonmonotonicity leads to a rather strong decrease of the
extrapolated exponent compared to the estimates of Ref. \cite{carl01}.

\begin{figure}[t]
\includegraphics[width=7.7cm]{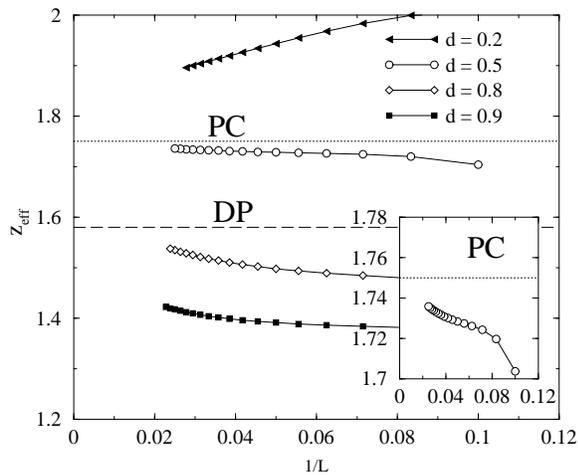}
\caption{
Plot of the effective exponent $z_{\rm eff}$ as a function of $1/L$ for
$d=0.2$, $0.5$, $0.8$, and $0.9$. Inset: blowup of the data for $d=0.5$.
}
\label{FIG05}
\end{figure}

\section{Results on the dynamical exponent}

{F}rom the ratio of the exponents $\beta/\nu_\perp$ and $\beta/\nu_\|$ one
can estimate the dynamical exponent $z = \nu_\|/\nu_\perp$.  Since both
$\beta/\nu_\perp$ and $\beta/\nu_\|$ are consistent with DP, also the
dynamical exponent $z$ agrees with the DP value $z_{\rm DP} =1.58$. It
is however instructive to show the results of an independent calculation
of $z$.  This quantity can be obtained from a finite size scaling analysis
of $\Delta$, the gap of the Master operator, which is the inverse of the
relaxation time of the system (see Ref. \cite{carl99} for details). As a
function of the system length $L$ the gap decays as $\Delta \sim L^{-z}$.

Figure \ref{FIG05} shows a plot of the effective exponent $z_{\rm eff}
= - \partial \ln \Delta/ \partial \ln L$ versus $1/L$. The calculations
are similar to those reported in Ref. \cite{carl01}, but now for longer
systems (up to $L = 46$ compared to $L=30$ of Ref. \cite{carl01}). The
critical point locations were obtained from Monte Carlo simulations, which
for this purpose are faster and more efficient than DMRG. Therefore we
concentrated our computational efforts on a single value of $p=p_c$
and could obtain results for longer systems.  As is clear from
Fig. \ref{FIG05}, the exponent $z_{\rm eff}$ is rather sensitive to the
value of the diffusion rate $d$.  As the estimates of $\beta/\nu_\perp$
and $\beta/\nu_\|$ are instead rather stable as a function of $d$ we
contribute this sensitivity to rather strong finite-size effects. Notice
that the finite $L$ corrections change sign from the weak to the strong
diffusion regime. The border value is around $d=0.5$ where $z_{\rm eff}$
has a very weak dependence on $L$. The data (see inset) run extremely
close to the PC value $z_{\rm PC} = 1.75$.  At higher diffusion rates
$d \approx 0.8--0.9$ the effective exponent $z_{\rm eff}$ for the range
of sizes investigated is much lower than $z_{\rm PC}$. At the strongest
diffusion investigated, extrapolations with different forms for the
correction to scaling terms as $1/L$ or $1/\sqrt{L}$ yield values in
the range $1.5 \lesssim z \lesssim 1.65$, which should be compared with
the DP value $z_{\rm DP} = 1.58$.  Current Monte Carlo estimates from
various authors \cite{hinr01,odor00,kock03} place the exponent $z$ in
the range $1.7-1.8$ and the calculations were mostly performed in the
weak diffusion regime.

\begin{figure}[t]
\includegraphics[width=7.5cm]{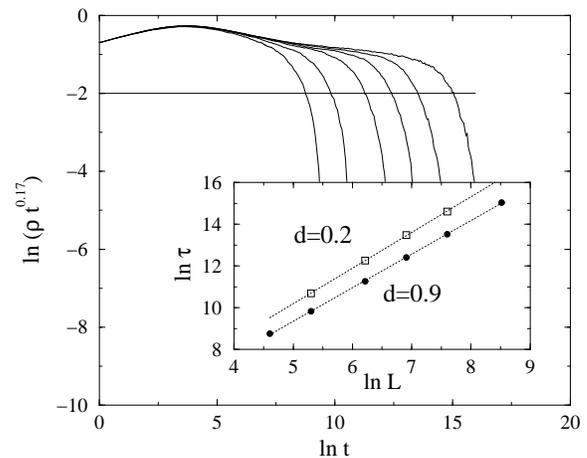}
\caption{
Scaled particle density for $d=0.9$ at the estimated critical point
$p=0.2335$ for $L=100$, $200$, $500$, $1000$, $2000$, and $5000$.
Inset: plot of $\ln \tau$ vs $\ln L$ for two values of the diffusion
constant. We estimate $z=1.61(3)$ for $d=0.9$ and $z=1.70(3)$ for $d=0.2$.
}
\label{FIG06}
\end{figure}

We also performed a series of Monte Carlo simulations to calculate
the exponent $z$ using finite-size scaling analysis.  At the critical
point and on a finite system the particle density decays as $\rho =
t^{-\beta/\nu_\|} f(t L^{-z})$, with $f$ a scaling function. For finite
$L$, $\rho$ follows a power-law decay up to a characteristic time $\tau$
after which it drops exponentially. One expects that $\tau$ scales as
$\tau \sim L^z$. To estimate $\tau$ we calculated $\rho$ for lattices
up to $L=5000$ and $t=10^7$ Monte Carlo time units. Figure \ref{FIG06}
shows a plot of $\ln (t^{\beta/\nu_\|} \rho)$ versus $\ln t$ for various
$L$ and $d=0.9$. The intersection of the data with a horizontal line at
$t^{\beta/\nu_\|} \rho = k$ (with $k$ a constant) provides an estimate
of $\tau$.  As we work in a region where the particle density is rather
low and fluctuations are large, and as the calculation of $z$ requires
very smooth data, we performed averages over a large number of samples
($> 10^3$).  For the calculation we used $\beta/\nu_\| = 0.17$, which
is the value determined above, and $k=-2$ (see Fig. \ref{FIG06}).
The inset shows a double-logarithmic plot of $\tau$ versus $L$ for
$d=0.2$ and $d=0.9$ at their critical points. In the former case we
restricted ourselves to $L=2000$ as the relevant times are typically
longer at weak than at stronger diffusion, as expected.  Notice that
in both cases the data are well fitted by straight lines yielding the
estimates $z=1.70(3)$ for $d=0.2$ and $z=1.61(3)$ for $d=0.9$, where
the latter value is consistent with the dynamical exponent of directed
percolation $z_{\rm DP} = 1.58$.  The results generally confirm the DMRG
findings according to which the dynamical exponent is generically smaller,
for finite $L$, at higher diffusivity.  We also notice that by varying
the value of $\beta/\nu_\|$ entering in the $y$ axis of Fig. \ref{FIG06}
one changes the estimate for $z$.  For instance, if we take $\beta/\nu_\|
= 0.20$, as calculated in Ref. \cite{kock03}, this leads to an increase
of $0.03$ in the estimated value for $z$.  The estimate of $z$ is rather
stable for changes in the constant $k$.

\begin{figure}[t]
\includegraphics[width=7.7cm]{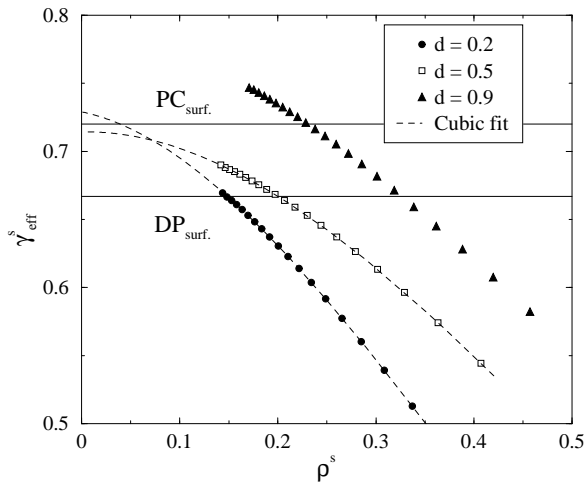}
\caption{Plots of $\gamma_{\rm eff}^s$ for three different values of
the diffusion constant.  Horizontal lines are the reference exponents
for the DP and PC universality classes.  }
\label{FIG07}
\end{figure}

\section{Results on surface densities}

Boundary quantities are easily accessible in DMRG techniques
\cite{carl97}, as one is basically forced to work with open boundary
conditions. Surface criticality in absorbing phase transitions has been
the subject of several studies in the past years both for models in the DP
\cite{laur97} and in the PC \cite{laur98} universality classes.  In the
latter case, it is known that there are two distinct surface exponents
depending on the type of boundary conditions applied \cite{laur98}.
The results of a DMRG calculation of the surface critical exponents for a
reaction-diffusion model in the PC class are presented in the Appendix.
Here, we report on the surface critical exponent calculations for the
PCPD, using the same types of boundary conditions as in the Appendix.

\begin{figure}[t]
\includegraphics[width=7.7cm]{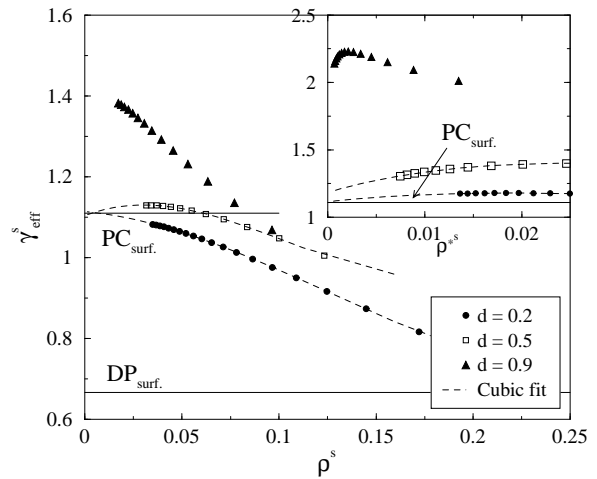}
\caption{Plots of $\gamma_{\rm eff}^s$ vs $\rho^s$ for different 
values of the diffusion constant and absorbing boundary conditions.
Inset: $\gamma_{\rm eff}^s$ vs $\rho^{* s}$ calculated from the 
surface pair density.
}
\label{FIG08}
\end{figure}

As in the calculation of the bulk particle density of the preceding
section we inject particles through the reaction $0 \to A$ at the boundary
site labeled by the position $i=1$ in order to induce a finite density of
particles in the system, and we measure the particle density $\rho^{\rm
s} (L)$ at the opposite boundary site $i=L$. Asymptotically for $L
\to \infty$, we expect $\rho^{\rm s} (L) \sim L^{-\beta^s/\nu_\perp}$,
where $\beta^{\rm s}$ is the order parameter surface exponent.  The two
different boundary conditions (BCs) applied at the site $i=L$ are: (a)
No particles are allowed to leave the system from the boundary site and
(b) particles may diffuse out of the system, i.e., the reaction $A \to
0$ (with rate $d$) is added at that site.  We refer to these as {\it
reflecting} and {\it absorbing} boundary conditions, respectively.

In Fig. \ref{FIG07} we plot the effective exponent $\gamma_{\rm eff}^s=
-\partial \ln \rho^{\rm s}(L)/\partial \ln L$ versus $\rho^s$ in the case
of reflecting BCs.  Horizontal lines show the ratio $\beta^s/\nu_\perp$
for DP ($=0.667$ \cite{laur97,carl99}) and PC ($=0.72$ \cite{laur98}). In
the DP case the different BCs produce the same critical exponent.
Effective exponents in this case grow monotonically, contrary to what is
found for bulk exponents.  Notice that a cubic fit yields a quite stable
estimate $\beta^s/\nu_\perp = 0.72(1)$ in the range $d \lesssim 0.5$,
a value actually consistent with the surface exponent for the PC class
(see Appendix).  Only at higher $d$ we observe some deviation from PC.
The fact that the extrapolated surface exponents vary with $d$, while
our current estimates for the bulk exponents are independent on $d$,
is an indication that the former are not yet the correct asymptotic ones.

Figure \ref{FIG08} shows $\gamma_{\rm eff}^s$ versus $\rho^s$ for the
case of absorbing boundary conditions. Again for weak diffusion the
exponent seems to extrapolate rather convincingly to values close to
the PC class ($\beta^s/\nu_\perp \approx 1.11$, see Appendix), while for
strong diffusion it increases to much larger values.  Also in this case
there is no clear signature of nonmonotonic behavior, except for the case
$d=0.5$ where the data for the largest systems pass through a maximum.

We also analyzed the effective exponent data from the pair density
$\rho^{* s}$ which are shown in the inset of Fig. \ref{FIG08} in the
case of absorbing BCs, and plotted as functions of $\rho^{* s}$.  In the
range $d \lesssim 0.5$ the data extrapolate close to the PC surface
exponent $\beta^s/\nu_\perp \approx 1.11$ as for the particle density.
At very strong diffusion ($d = 0.9$) the surface effective exponent
shows a nonmonotonic behavior with a maximum around $\gamma_{\rm
eff}^s \approx 2.2$.  Notice that particle and pair exponents in this
case are rather far apart from each other and it is quite hard to find
a common extrapolation value. We would expect for $\gamma_{\rm eff}^s$
a similar behavior as for the bulk exponents, i.e., an increase followed
by a decrease towards the asymptotic value.  We suspect that in the
present surface exponent calculation the decreasing side has barely been
reached. So we tend to distrust the extrapolation as estimates of the
genuine asymptotic behavior.  They rather provide some insight on the
preasymptotic region and actually point to a similarity with PC surface
exponents at weak diffusion.

\section{Discussion}
\label{sec:discussion}

To conclude, by combining Monte Carlo and DMRG calculations we
analyzed the critical properties of the pair contact process with
diffusion. This model has been the subject of increasing attention in
recent years. Although the debate around it has not yet been settled,
the main belief is that the PCPD belongs to a novel universality class
which differs from the known DP and PC classes.

In our opinion, however, the most plausible scenario for the PCPD is
that it ultimately falls into the DP universality class. The asymptotic
behavior is, however, masked by rather strong finite size and time effects,
characterized by small correction-to-scaling exponents, as our Monte
Carlo simulations for the decays of the particle density $\rho$, the
pair density $\rho^*$, and the ratio $\rho / \rho^*$ have demonstrated.

The exponents $\beta/\nu_\|$ and $\beta/\nu_\perp$ extrapolated both from
$\rho$ and $\rho^*$ appear to be stable as functions of the diffusion
constant $d$ and actually consistent with the DP class values.  The data
show a nonmonotonic behavior both in time and system size, which in our
opinion points to a crossover phenomenon between two competing types of
critical behavior.  The surface exponents, which we also investigated,
turned out to be instead rather sensitive to the value of $d$; a sign,
in our opinion, that the extrapolated values are probably not the true
asymptotic ones. Interestingly enough, particularly at weak diffusion,
the extrapolated values are rather stable and consistent with those
expected for the PC class.

In early numerical studies of the PCPD \cite{carl01,hinr01,odor00},
restricted to shorter simulation times and system lengths compared
to those considered here, several quantities as $\beta/\nu_\perp$,
$\beta/\nu_\|$, and $\nu_\|/\nu_\perp$ were found to be quite consistent
with the PC class values.  It is now generally agreed that the PCPD
does not belong to the PC universality class, as more extensive
simulations performed by several groups have shown convincingly
\cite{noh01,park02,dick02,hinr02a,kock03}. Still one would like to
understand if the observed similarity with the PC exponent is purely
fortuitous or if there is some deeper reason for it. In our opinion the
evidence given above that also the surface exponents extrapolate towards
PC values in an intermediate regime strongly suggests that there is a
genuine nonasymptotic PC-like regime, with a crossover to DP behavior
at longer time scales.

A prototype model in the PC class is the branching annihilating random
walk with even offsprings (BARWe), defined by the reactions $A \to 3A$,
$2A \to 0$ plus diffusion \cite{hinr00,marr96}, which differs from the
PCPD only for the reaction which creates particles. We argue that the
early stages of the critical dynamics, when the system has a rather
high particle density, are dominated by the annihilation process $2 A
\to 0$, so that the substitution of the BARWe reaction $A \to 3A$ with
that of the PCPD $2A \to 3A$ may result in a very weak perturbation of
the system. Therefore a transient PC-like regime may be observed for $t
\lesssim \tau_{\rm c}$, where $\tau_{\rm c}$ is some crossover time.
This argument may help to explain features observed in the PCPD, and
should be equally valid for other models where the annihilation is of
the type $2A \to 0$ and with different creation rules $n A \to (n+k)
A$ with $n \geq 2$, $k > 0$; for such systems we expect a transient PC
regime as well.

The study of reaction-diffusion systems where the annihilation and
creation reactions involve $n \geq 2$ particles has recently drawn
some attention \cite{kock03,park02b,odor02}. In particular, we mention
here the two cases recently considered by \'Odor \cite{odor02} (i)
$3 A \to 5 A$, $2A \to 0$ and (ii) $4 A \to 5 A$, $2A \to 0$. In model
(i) he estimates $\beta/\nu_\| \approx 0.28$ (consistent with PC) for
small diffusion rates and  $\beta/\nu_\| \approx 0.24$ at stronger
diffusion. Invoking some logarithmic corrections he claims that all
values extrapolate to $\beta/\nu_\| \approx 0.22$ \cite{odor02}. In
case (ii) the estimate is $\beta/\nu_\| \approx 0.28$ both at high and
low diffusions \cite{odor02}, again consistent with the PC class value.
The above observations suggest that these types of systems follow closely
a critical behavior as described here for the PCPD, and it is thus
plausible that they fall for sufficiently long times into the DP class.
However, it may turn out to be quite difficult to show this numerically,
as we expect that increasing the number of particles involved in the
creation and annihilation reactions will lead to models even harder to
simulate and analyze than the PCPD.

Very recently Kockelkoren and Chat\'e analyzed a similar set of models
\cite{kock03}. In their formulation the fermionic constraint of only one
particle per site is released. All the reactions of the type $n A \to
(n+k)A$ and $2 A \to 0$ with $n > 2$ were found to belong to the DP class.
Surprisingly, in all those models the convergence to DP exponents seems to
be quite fast (at least for $\beta/\nu_\|$) and not plagued by the strong
corrections found in the fermionic models.  It would be interesting to
study the same models at different values of the diffusion constant, as
in the PCPD the onset of crossover behavior is quite strongly influenced
by the value of $d$.  

We are grateful to J.D. Noh, H. Hinrichsen, M. den Nijs, and F. van Wijland
for useful discussions.

\begin{figure}[t]
\includegraphics[width=7.7cm]{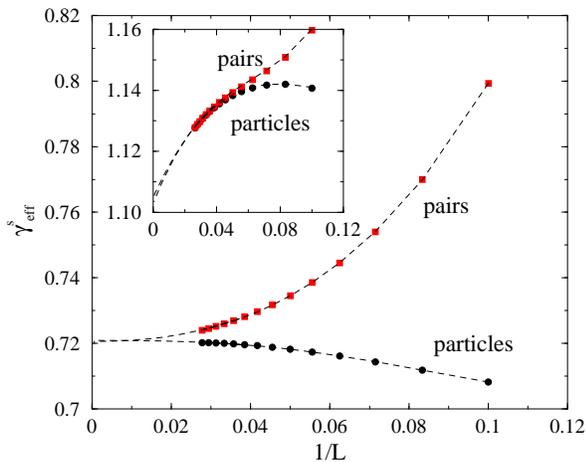}
\caption{
Plot of the surface effective exponent $\gamma^{\rm s}_{\rm eff}$ vs
$1/L$ for the parity conserving process $A \to 3A$ and $2A \to 0$ in 
the case of reflecting boundary conditions.  
Inset: $\gamma^{\rm s}_{\rm eff}$ vs $1/L$ for absorbing boundary 
conditions.  Dashed lines are extrapolated curves through the DMRG 
data. Extrapolated values are in good agreement with Monte Carlo 
simulations results of Ref. \cite{laur98} (see text).
}
\label{FIG09}
\end{figure}

\appendix*
\section{Surface critical behavior in the parity conserving process 
2A $\to$ 0, A $\to$ 3A}

We present here some results on the surface critical behavior of the
parity conserving process defined by the reactions $2A \to 0$, $A \to 3A$
and with single particle diffusion. We show how DMRG produces accurate
surface critical exponents for this model, which are in good
agreement with former Monte Carlo simulation results.
For the single particle diffusion and pair annihilation we used the
same rates as in Eq. (\ref{eq1}), while we assign a rate $(1-p)(1-d)$
to the reaction $0A0 \to AAA$. We restrict ourselves to a single
value of the diffusion constant $d=0.5$. 

We first estimated the critical point at $p=p_c \approx 0.577(2)$ by
means of Monte Carlo simulations using a standard approach. As mentioned
above, for surface universality in PC processes there are two possible
types of boundary conditions leading to two distinct surface exponents
\cite{laur98}.  In the first case, the system is truncated at one edge and
no particles are allowed to cross the boundary site; we refer to this as
{\it reflecting} boundary conditions. In the second case, particles are
allowed to drop from the boundary. We implemented this type of boundary
condition adding the boundary reaction $A \to 0$ (with rate $d$) which
mimics the diffusion of particles out of the system. We refer to this
implementation as {\it absorbing} boundary conditions.

Figure \ref{FIG09} shows the effective surface exponent $\gamma^{\rm
s}_{\rm eff}$ versus $1/L$ in the case of reflecting boundary
conditions, calculated both from the particle (circles) and pair
(squares) densities.  The same quantities are plotted in the inset
in the case of absorbing boundary conditions.  Notice that indeed the
results confirm the existence of two distinct sets of surface exponents
and that the data from pairs and particles merge for sufficiently long
chains, indicating that both quantities decay with the same exponent.
Our estimates $\beta^s/\nu_\perp \approx 0.720(2)$ in the former case
and $\beta^s/\nu_\perp \approx 1.10(1)$ in the latter are obtained
from a polynomial extrapolation in $1/L$. As the finite-size effects
are rather small (see Fig. \ref{FIG09}), the extrapolated values are
not very sensitive to the type of correction to scaling term used in
the extrapolation.

The Monte Carlo simulation results \cite{laur98} for the critical
exponents are $\beta^{\rm s} = 1.34(2)$ and $\beta^{\rm s} = 2.04(2)$,
for inactive and active boundary conditions, respectively. Combining
these results with the PC class correlation length exponent $\nu_\perp
= 1.83(3)$ \cite{hinr00}, one finds $\beta^s/\nu_\perp = 0.73(1)$
(reflecting BCs) $\beta^s/\nu_\perp = 1.11(1)$ (absorbing BCs), in very
good agreement with the DMRG calculations.


\begin{references}
\bibitem{ratio}
	One can show analytically that this quantity tends to a constant 
	in the contact process, which belongs to the DP universality class.
\end{references}
\end{document}